\documentclass[12pt]{article}

\usepackage{graphicx} 

\def\beq{\begin{equation}}
\def\eeq{\end{equation}}
\def\beqa{\begin{eqnarray}}
\def\eeqa{\end{eqnarray}}

\def\-{\hphantom{-}}

\def\s2{\frac{1}{\sqrt2}}

\def\beq{\begin{equation}}
\def\eeq{\end{equation}}
\def\beqa{\begin{eqnarray}}
\def\eeqa{\end{eqnarray}}
\def\ba{\begin{array}}
\def\ea{\end{array}}

\def\IF{\relax{\rm I\kern-.18em F}}
\def\II{\relax{\rm I\kern-.18em I}}
\def\IP{\relax{\rm I\kern-.18em P}}
\def\IC{\relax\hbox{\kern.25em$\inbar\kern-.3em{\rm C}}}
\def\IR{\relax{\rm I\kern-.18em R}}

\def\Dsl{\,\raise.15ex\hbox{/}\mkern-13.5mu D} 
\def\IZ{Z\kern-.4em  Z}

\begin{document}
%

\begin{center}
\begin{large}
{\bf  Non Hamiltonian Chaos from Nambu Dynamics of Surfaces\footnote{Published in Chaos Theory: Modeling, Simulation and Applications,\\C.H. Skiadas, I. Dimotikalis and C. Skiadas (Eds), \\World Scientific Publishing Co, pp. 110-119.}   }

\end{large}

\vskip1truecm

{Minos Axenides}
\end{center}

%
%
%

%
%

\begin{center}
  Institute of Nuclear Physics, NCSR Demokritos,
  15310 Agia Paraskevi, Attiki, Greece\\
  (E-mail:  axenides@inp.demokritos.gr)
\end{center}

\begin{abstract}
We discuss recent work with E.Floratos (JHEP 1004:036,2010) on Nambu Dynamics of
 Intersecting Surfaces  underlying  Dissipative Chaos in $R^{3}$. 
We present our argument for the  well studied  Lorenz and  R\"{o}ssler strange attractors. 
We implement a flow decomposition to their equations of motion.
Their  volume preserving part preserves in time a
family of two intersecting surfaces, the so called { \em Nambu Hamiltonians}.
For dynamical systems with linear dissipative sector such as the Lorenz system, they are specified in terms
 of Intersecting Quadratic Surfaces. For the case of  the  R\"{o}ssler system, with nonlinear dissipative part,  they are given in terms of a Helicoid intersected by a Cylinder.
In each case they foliate the entire phase space and get  deformed  by
Dissipation , the irrotational component to their flow.
It is given by the gradient of a surface in $R^{3}$ specified in terms of a
scalar function. All three intersecting  surfaces reproduce completely the dynamics 
of each strange attractor.
\end{abstract}

\section{Introduction}
Dissipative dynamical systems, with a low dimensional phase space, present
an important class of simple non-linear physical systems
with intrinsic complex behavior (homoclinic bifurcations, period doubling,
onset of chaos, turbulence), which generated
intense experimental, theoretical
and numerical work in the last few decades
\cite{one,turbB}.

Recently \cite{AF1} we have reexamined dissipative dynamical
systems with a 3-dimensional phase space from the perspective of Nambu-Hamiltonian
Mechanics(NHM) \cite{Nam,Tak}. The latter  represents a generalization of
Classical Hamiltonian Mechanics, mostly appropriate for the study of odd-dimensional
phase-space volume preserving flows( Liouville's theorem). As such and 
in order to make it directly applicable
to the dynamics of dissipative systems in $R^{3}$ we must associate it to a volume 
preserving dynamics sector. 
We have done so by introducing a flow decomposition to 
their equations of motion and therefore isolate 
in their flow vector field its rotational(solenoidal) part. 
It is manifestly volume preserving,  non-dissipative  and hence directly describable 
in terms of the intersecting surfaces of NHM. Along with the remaing irrotational component to the flow 
the decomposition fully
recovers  the dissipative dynamics of the system in question, presenting itself as an 
equivalent formulation of the odd-dimensional dynamical system.

We have applied the above recipe to the  famous examples of Lorenz\cite{Lor}
and R\"{o}ssler
\cite{Ross}
 chaotic attractors
 which represent the prototype models for the onset of turbulence.\cite{turbB}.

     In sect. 2  we start off with a discussion of flow decompostion for the most elementary and
     familiar of  all dissipative systems in $R^{2}$ without chaotic behavior : 
     the dissipative harmonic oscillator (DHO). 
    In this case the existence of a Hamiltonian formalism identifies  for its "closed" nondissipative sector 
    the harmonic oscilator (HO) with a well defined integrable classical and quantum 
    evolution. 
     
     The  presence of chaotic flows for dissipative systems in $R^{3}$ is argued, by analogy, 
      to nescessitate intersecting Nambu Surfaces which define rerspectively  "closed" physical systems 
      with integrable $R^{3}$ periodic orbits and simultaneously a well
     defined classical and quantum behaviour.  
     
    In sect. 3 we apply the framework for  the cases of the
       Lorenz and R\"{o}ssler strange attractors. 
      We isolate their non-dissipative sector parametrized by
       two intersecting surfaces: a cylinder and a paraboid for the Lorenz attractor 
     as well as a helicoid with a cylinder for the  R\"{o}ssler system.
       They  account, amazingly, for the double scroll topology of the full " butterfly"
      Lorenz  attractor. and the single scroll topology for the   R\"{o}ssler case.
     We end our presentation with Conclusions and open problems.

\section{Flow Decomposition in Dissipative Systems: $R^{2}$ versus $R^{3}$ }


Dissipation is a necessary condition for dynamical systems to exhibit chaos. Yet it is not suficient. This is enunciated through a powerful No-go theorem . Indeed the Poincare-Bendixon alows for
only fixed points and limits cycles in two dimensions.  Chaotic Flows need space to  emerge. At the 
minimal level they emerge with dissipative systems in $R^{3}$ . They are typically associated with
Strange Attractors such as the famous  ones of Lorenz and R\"{o}ssler.  They belong to a large class of dynamical systems whose dynamics is governed by continuous set of 1st order ordinary differental 
flow equations 
\beq
\dot{ \vec{x} } \ =\  \vec{v} ( \vec{x} (t), t , \lambda)
\eeq
where $\vec{v} $ is a velocity field flow with $\lambda$ some control external parameter. Their
phase space dynamics, depending on whether they exchange energy with their environment or not, 
 can be either {\em open-dissipative} or {\em closed (conservative-Hamiltonian) }. 
This is reflected on their velocity flow field being  divergenceless or not.
Let us  see all these issues for the simple case of the Dissipative Harmonic Oscillator (DHO) whose phase space equations of motion are well known to be for $ x_{1}=q $ and $x_{2}=p$:

\beqa
\dot{q} \ &=& \ p \ + \ \alpha q  \nonumber  \\
\dot{p} \ &=& \ -q + \beta p 
\eeqa   
The velocity flow is given by $ \nabla \dot v_{tot} = (\alpha + \beta) \neq 0 $  for appropriate values 
$ \alpha \neq -\beta \neq 0 $ with  a net outflow 
or inflow of energy depending on the sign of  $\nabla \dot v_{tot}$ (inflow $ < 0 $  or outflow $> 0 $).
Nevertheless we split it into a  nondissipative component 
$ v_{ND} = ( p, -q )$ with zero flow  $ \nabla \dot v_{ND} = 0 $.  
Its dissipative part is given
by  $ v_{D} = \alpha q + \beta p $ with  its  flow being simply the total flow of the HO
$ \nabla v_{D} = \alpha + \beta $.  The dynamics of the HO in one dimension  can take the
form of a 2nd order differential equation
\beq
\ddot{q}  +  (\alpha + \beta ) \dot{q} + (1+\alpha \beta ) q = 0
\eeq
The damping factor is $ \gamma = \alpha+\beta $ and the effective natural frequancy which is
dissipation induced is given by  $\omega_{eff}= 1+\alpha\beta$ .

The existence of the Harmonic Oscillator Hamiltonian   $ H = \frac{1}{2} ( p^{2} + q^{2} ) $ 
along with a  Dissipation function $D=\frac{1}{2} ( \alpha q^{2}+\beta p^{2} )$ reproduce the eqs. of motion of (2.2) in a more compact form
\beq
\dot{x}^{i} \ = \  \epsilon^{ij} ( \partial_{j} H + \partial_{j} D )
\eeq
where $i,j=1,2$  and $\epsilon^{12}=-\epsilon^{21}=1$ as usual.
The non-dissipative sector of the HO is given by 
   
\beqa
\dot{q} \ &=& \  \frac{\partial H}{\partial p} \ = \  p \nonumber  \\
\dot{p} \ &=&  \ - \frac{\partial H} {\partial q} \ = \ - q  
\eeqa

{\em Firstly} it identifies a circular {\em periodic } orbit, the famous HO  as the {\em integrable progenitor} 
of the transient DHO . 
This is a well defined closed  physical system with a most familiar 
quantum behavior,
the Quantum Harmonic Oscillator.

{\em Secondly} the linear dissipation of the DHO a transient  spirals it inwards to a 
fixed point of zero energy. The
harmonic oscillator {\em localizes } in effect such a  dissipative evolution .

{\em Lastly}  the rate at which the energy of the harmonic oscillator loses its energy as it spirals in 
 depends on its damping strength. It is the  {\em Quality factor} which is easily computed to be
  
\beq
 Q \ = \ \frac{\omega_{eff}}{\gamma} \ = \  \frac{ 1 + \alpha \beta}{ \alpha + \beta}
\eeq   
All in all  Flow decomposition for the transient DHO has three immediate implications for the DHO:
1. Existence of  Integrable Harmonic oscillator associated with a periodic  orbit 
2.  Localization of the DHO evolution and asymptotic fixed point state
3.  Quality factor for DHO as a measure of its energy loss rate and in effect of the damping strenth.

We will proceed now to examine whether it is feasible to implement this methodology for the
case of  Strange attractors  which  possess chaotic flows in $R^{3}$.

Nambu-Hamiltonian mechanics is a specific generalization of classical
Hamiltonian mechanics, where the invariance group of canonical
symplectic transformations of the
Hamiltonian evolution equations in 2n dimensional phase space is extended to the
more general
volume preserving transformation group $ SDiff({\cal M}) $ with an arbitrary phase space
manifold {\cal M} of any dimension $d=\mbox{dim}({\cal M}) $ .

In  \cite{AF1} we 
work with the case of a three dimensional flat phase space
manifold. Nevertheless our results could be generalized to curved manifolds of any
dimension\cite{Tak,AF}.

Nambu-Hamiltonian mechanics of a particular dynamical system in $R^{3}$ is defined once
two scalar functions $H_{i}\in C^{\infty}(R^{3}), i=1,2 $, the generalized
Hamiltonians \cite{Nam,Tak} are provided.
The evolution equations are:
\beq
\dot{x}^{i} \ = \  \{ x^{i} , H_{1} , H_{2} \} \ \ \ \ \ \ \ \  i=1,2,3
\eeq
where the Nambu 3-bracket, a generalization of Poisson bracket in
Hamiltonian mechanics, is defined as
\beq
\{ f, g, h \} \ = \ \epsilon^{ijk} \partial^{i}f \partial^{j}g \partial^{k}h
\ \ \ \ \ \ i,j,k=1,2,3 \ \ \ \ \ \ \forall f,g,h \in C^{\infty}( R^{3})
\eeq

Any local coordinate transformation
\beq
x^{i} \ \rightarrow \ y^{i} \ = \ y^{i}(x) \ \ \ \ \ \ \ \ \ i=1,2,3
\eeq
which preserves the volume of phase space
\beq
\mbox{det} \left( \frac{\partial y^{i}}{\partial x^{j} } \right) \ = \ 1 \ \ \ \ \
 \forall \ x=(x^{1},x^{2},x^{3}) \in R^{3}
\eeq
leaves invariant the 3-bracket and therefor it is a symmetry of Nambu-Mechanics.
Except for the linearity and antisymmetry of the bracket with respect to all of
its arguments it also
satisfies an important identity, the so called "Fundamental identity"[FI] \cite{AF1,AF}.
 
The evolution eq.(2.7) has a flow vector field:
\beq
v^{i}(x) \ \ = \ \ \epsilon^{ijk}\partial^{j} H_{1} \partial^{k}H_{2} \ \ \ \
\ \ \ \ \ i,j,k=1,2,3
\eeq
which is volume preserving
\beq
\partial^{i} v ^{i} \ = \ 0
\eeq
The reverse is also true. 
We will name the phase-space volume preserving flows "Non-dissipative"
while the non-conserving ones "Dissipative" ($\partial^{i} v^{i}(x) > 0(< 0))$.

The flow equations of a general dissipative system may take the general vector  form

\beq
\dot{\vec{x}} \ \ = \ \ \vec{\nabla }H_{1} \times \vec{\nabla }H_{2}
+ \vec{\nabla }D
\eeq
We notice that given a pair of functions $H_{1}, H_{2}$ such that
\beq
\vec{v}_{ND} \ \ = \ \  \vec{\nabla }H_{1} \times \vec{\nabla }H_{2}
\eeq
any transformation of $H_{1},H_{2}$
\beq
H_{i} \rightarrow H_{i}^{\prime }(H_{1}, H_{2}) \ \ \ \ \ i=1,2
\eeq
with unit Jacobian
\beq
\mbox{det} \left( \frac{\partial H_{i}^{\prime }}{\partial H_{j}} \right) \ = \ 1
\eeq
gives also
\beq
\vec{v}_{ND} \ = \ \vec{\nabla }H_{1}^{\prime } \times \vec{\nabla }H_{2}
^{\prime }
\eeq

In ref.\cite{AF} we reduced the evolution equation of the form (2.32)
Nambu Mechanics in Hamiltonian-Poisson form as follows:
\beq
\dot{x}^{i} \ = \ \{ x^{i} , H_{1} , \}_{H_{2}}
\eeq
where the induced Poisson bracket
\beq
\{ f , g \}_{H_{2}} \ = \ \epsilon^{ijk}\partial^{i}f \partial^{j}g \partial^{k} H_{2}
\eeq
satisfies all the required properties like linearity, antisymmetry
and the Jacobi identity.

In the following sections  we are going to present a detailed investigation of the
Lorenz and R\"{o}ssler attractors from the point of view of Dissipative Nambu-Hamiltonian Dynamics.

\section{ The Lorenz Attractor from Dissipative Dynamics of Intersecting Quadratic Surfaces .}

The Lorenz model was invented as a three Fourier mode truncation
of the
basic eqs. for heat convection in fluids in Reyleigh-Benard type of experiments
\cite{Tab}
The time evolution eqns. in  the space of three Fourier modes $ x,y,z $ which we
identify as phase-space are :
\beqa
\dot{x} \ &=& \ \sigma ( y - x )    \nonumber  \\
\dot{y} \ &=& \ x(r-z) - y        \nonumber   \\
\dot{z} \ &=& \ xy - bz
\eeqa
where $\sigma $ is the Prandtl number, r is the relative Reynolds number
and b the geometric aspect ratio .

The standard values
for $ \sigma $, b are  $ \sigma=10, b = \frac{8}{3} $ with r taking values
in $ 1 \leq r < \infty $. There are dramatic changes of the
system as r passes through various critical values which follow the
change of stability character of the three critical points of the
system $P_{1} $ : $ x=y=z=0 , P_{\pm }: x=y=\pm \sqrt{b(r-1)}, z=r-1 $

Lorenz discovered the non-periodic deterministic
chaotic orbit for the value $r=28 $, which is today identified as a
Strange Attractor with a Hausdorff dimension of ($ d=2.06 $)\cite{FOY}. Standard
reference for an exhaustive numerical investigation of the Lorenz system
is the book by Sparrow \cite{Spa}.

There
have been various attempts made to localize the Lorenz attractor, by convex surfaces,
in order to get information about Hausdorff dimensions \cite{FOY}
and other characteristics
\cite{gattL}.

We will proceed to exhibit Localization of the full Lorenz attractor from the Nambu surfaces which
we will determine shortly. At the Quantum level the existence of an attracting ellipsoid in a matrix
model formulation of the Lorenz system is manifest \cite{AF1} as one gets attracting ellipsoids in
higher dimensional phase spaces.

We now proceed to describe the Lorenz system in the framework  of section 2.
The flow vector
field $ \vec{v} $ is analyzed into its dissipative and non-dissipative
parts as follows:
\beq
\vec{v}_{D} \ = \ ( -\sigma x , -y , -b z ) \ = \ \vec{\nabla } D
\eeq
with the "Dissipation" function
\beq
D\ = \ - \frac{1}{2} \ ( \sigma x^{2} \ + \ y^{2} \ + \ b z^{2})
\eeq
and
\beq
\vec{v}_{ND} \ = \ ( \sigma y, x(r-z), xy) \ = \ (0 , y , z-r) \times
(-x , 0 , \sigma )
\eeq
The two {\em Hamiltonians} or Clebsch-Monge potentials
$ H_{1}, H_{2} $ are determined by
\beq
\vec{\nabla} H_{1} \times  \vec{\nabla} H_{2} \ = \ \vec{v}_{ND}
\eeq
or equivalently
\beq
H_{1} \ = \ \frac{1}{2} [ y^{2} + ( z- r )^{2} ]
\eeq
and
\beq
H_{2}\ = \ \sigma z \ - \ \frac{x^{2}}{2}
\eeq
The Lorenz system (20) can thus be written in the equivalent form in terms  of  $\vec{r}=(x,y,z) $:
\beq
\dot{\vec{r}} \ = \  \vec{\nabla} H_{1} \times  \vec{\nabla} H_{2}
\ + \ \vec{\nabla }D
\eeq
In the Non-Dissipative part(ND) of the dynamical system
\beq
\dot{\vec{r}} \ = \  \vec{\nabla} H_{1} \times  \vec{\nabla} H_{2}
\eeq
the Hamiltonians $ H_{1}, H_{2} $ are conserved and their intersection
defines the ND orbit.
Moreover if we get the reduced Poisson structure
(sect.2) from $ H_{2} $ we obtain the 2-dim phase space
$ \Sigma_{2}$  to be the family of parabolic cylinders
with symmetry axis the y-axis:
\beq
H_{2}=\mbox{constant} = H_{2}(\vec{r_{o}})
\eeq
 $ \Sigma_{2}$ is thus given by
 \beq
 z\ = \ z_{o} + \frac{x^{2} - x_{o}^{2}}{2\sigma}
 \eeq
 with $ x_{o},z_{o}$ the initial condition for $ x,z $.
 The induced Poisson algebra (rel.19) is given by
 \beqa
 \{ x , y \}_{H_{2}} \ &=& \ \partial_{z}H_{2} \ = \sigma \nonumber \\
 \{ y , z \}_{H_{2}} \ &=& \ \partial_{x}H_{2} \ = -x \nonumber \\
 \{ z , x \}_{H_{2}} \ &=& 0
 \eeqa

The dynamics on the 2d-phase space $ \Sigma_{2} $ is given by $H_{1} $
\beqa
\dot{x} \ &=& \ \{ x , H_{1} \}_{H_{2}} \ \ \ \    \nonumber \\
\dot{y} \ &=& \  \ \{ y , H_{1} \}_{H_{2}}            \nonumber \\
\dot{z} \ &=& \ \  \{ z, H_{1} \}_{H_{2}}
\eeqa
and $H_{1}$ is an anharmonic oscillator Hamiltonian with $( x / \sigma, y )$
conjugate canonical variables.  Using rel.(25-26) we get on $ \Sigma_{2}$:
\beq
H_{1} \ = \ \frac{1}{2\sigma^{2}} \ [ y^{2} \ + \ \frac{1}{2} \ ( x^{2} \ - \ a^{2})^{2}]
\eeq
with
\beq
a^{2} \ = \ x_{o}^{2} \ - \ 2 \sigma ( z_{o} \ - \ r ) \ = \
- 2 H_{2} \ + \ 2 \sigma r
\eeq
where  $ \frac{1}{\sigma^{2}} $ plays the role of the mass.
Depending on the initial conditions we may have a single well
$(a^{2} \leq 0 , H_{2} \geq \sigma r )$ or a double well
potential ( $  a^{2}>0 , H_{2} < \sigma r $ ) respectively.
The trajectories , the intersections
of the two cylinders , $H_{1}$ and $H_{2}$ with orthogonal symmetry axes $(x,y)$ may
either have one lobe left/right or may be running from the right to the left lobe. This
is reminiscent of the topology structure of the orbits of the
Lorenz chaotic attractor.

The full Lorenz system does not conserve
$H_{1}, H_{2} $ and there is a random motion of the two surfaces against each other. Their intersection
is time varying.
In effect at every moment the system jumps from periodic to periodic orbit
of the non-dissipative sector.
Moreover the motion of the non-dissipative system around the two lobes,
either left or right, can now jump from time to time from one lobe to the other.

\section{ The R\"{o}ssler Attractor from  Dissipative Dynamics of
a Cylinder Intersecting with an Helicoid}
R\"{o}ssler introduced a simpler than Lorenz's nonlinear ODE system with
a 3d- phase space, in order to study in more detail the characteristics of
chaos, which is motivated by simple chemical reactions \cite{Ross}.

The R\"{o}ssler system is given by the evolution eqns:
\beqa
\dot{x} \ &=& \ - y - z \nonumber \\
\dot{y} \ &=& \ x + a y \nonumber \\
\dot{z} \ &=& \ b + z(x-c)
\eeqa
with parameters $ a, b, c $ usually taking standard values
$ a=b=0.2, c=5.1 $ or
$ a=b=0.1, c=14 $ for the appearance of the chaotic attractor. 

We turn now to the study of the R\"{o}ssler system as a Dissipative
Nambu-Hamiltonian dynamical system.

The key difference with the Lorenz attractor is that the dynamics of
the system is simpler. Chaos appears as random jumps outwards and inwards
the single lob attractor.

In order to get the three scalars,
the two generalized Hamiltonians $ H_{1} , H_{2} $ which are conserved and
characterize the non-dissipative part and D the  dissipation term :
\beq
\dot{\vec{r}}  \ = \ \vec{\nabla} H_{1} \times \vec{\nabla}H_{2} \ + \
\vec{\nabla} D
\eeq
we checked after some guess work, that we must subtract and add a new
term in the first equation. Indeed we find for the two parts
\beq
\vec{v}_{ND} \ = \ (-y - z - \frac{z^{2}}{2}, x , b )
\eeq
\beq
\vec{v}_{D} \ = \ ( \frac{z^{2}}{2}, a y , z (x-c) )
\eeq
satisfying accordingly $\vec{\nabla }\cdot \vec{v}_{ND} \ = 0$ and $\vec{\nabla} \times \vec{v}_{D} \ = \ 0$

We must determine $ H_{1} , H_{2} $ and D such that
\beq
\vec{\nabla} H_{1} \times \vec{\nabla}H_{2}\ = \ \vec{v}_{ND}
\eeq
and
\beq
\vec{\nabla } D \ = \ \vec{v}_{D}
\eeq
For D we find easily
\beq
D \ = \  \frac{1}{2} \ [ a y^{2} + (x-c)z^{2}]
\eeq
To get $ H_{1}, H_{2} $ we must integrate first the Non-dissipative system:
\beqa
\dot{x} \ &=& \ -y -z - \frac{z^{2}}{2} \ \ \ \ \ \nonumber \\
\dot{y} \ &=& \ x \ \ \ \ \ \ \ \nonumber \\
\dot{z} \ &=& \ b
\eeqa
The general solution is :
\beq
x(t) \ = \ - b ( 1 + z(t)) \ + \ ( x_{o} + b(1+z_{o}) )\mbox{ cos(t)}
- ( y_{o} + z_{o} + \frac{z_{o}^{2}}{2} - b^{2} )\mbox{ sin (t)}
\eeq
\beq
y(t) \ = \ b^{2} - z(t) - \frac{z^{2}(t)}{2} +  ( y_{o} + z_{o} +
\frac{z^{2}_{o}}{2} - b^{2})\mbox{ cos(t)} \ + \ ( x_{o} + b + b z_{o} )\mbox{ sin (t)}
\eeq
\beq
z(t) \ = \ b t + z_{o}
\eeq
To uncover $ H_{1} , H_{2} $ we introduce the complex variable
\beq
w(t) \ = \ w_{1}(t) + i w_{2}(t)
\eeq
with
\beqa
w_{1}(t) \ &=& \ x(t) \ + \ b ( 1 + z(t) )  \nonumber \\
w_{2}(t) \ &=& \ y(t) + z(t) + \frac{z^{2}(t)}{2} - b^{2}
\eeqa
We obtain
\beq
w(t) \ = \ w_{o} \cdot e^{i t}
\eeq
with
\beq
w_{o} \ \equiv  \ w(t=0)
\eeq
We see that there are two constants of motion, the first one being :
\beq
\mid w(t) \mid  \ = \ \mid w_{o} \mid
\eeq
and we define correspondingly,
\beq
H_{1} \ = \ \frac{1}{2} \mid w(t) \mid ^{2} \ = \ \frac{1}{2} \
( x + b (1 + z) )^{2} \ +
\ \frac{1}{2} \ ( y + z + \frac{z^{2}}{2} -b^{2})^{2}
\eeq
The second integral of motion is obtained through the phase
\beq
w(t) \ = \ \mid  w_{o} \mid \cdot e^{\imath \varphi _{o}} \cdot e^{\imath t} \ =
\ \mid w_{o}\mid e^{\imath \varphi(t)}
\eeq
or from (4.14)
\beq
\varphi (t) \ - \ \frac{z}{b} \  = \ \varphi_{o} - \frac{z_{o}}{b}
\eeq
and we define appropriately the second constant surface $ H_{2} $
\beq
H_{2} \ = \ b \ \ \mbox{arctg} \frac{ y + z + \frac{z^{2}}{2} - b^{2}}
{1 + b(1+z)} \ \ - \ z
\eeq
We easily check that rel.(41) is satisfied:
\beq
\vec{\nabla} H_{1} \times \vec{\nabla}H_{2}\ = \ ( -y -z - \frac{z^{2}}{2} , x , b )
\eeq

The family of surfaces $ H_{1}$ and $ H_{2} $ are a quadratic deformation
of a cylinder and respectively a quadratic deformation of a right helicoid.
Their intersection is the trajectory (43-44).

\section{ Conclusions-Open Problems}
The main result of our present work is the demonstration of Dissipative
Nambu-Hamiltonian mechanics of intersecting surfaces
as the conceptual framework that underlies strange chaotic attractors
both in their classical as well as quantum-noncommutative incarnation.
  It reproduces
 the familiar and well studied attractor dynamics of Lorenz and R\"{o}ssler  in a very
 intuitive manner accounting  of their gross topological aspects 
 (double lobe Butterfly for the Lorenz system )
 or single lobe for
the R\"{o}ssler attractor.

 Quantum Nambu Dynamics of Surfaces  \cite{AF1,AF,AFN}, 
 raises also the issue of possible existence of Quantum Strange Attractors.
 Their Quantum behavior was  built systematically
 through  { \em fuzzifying } the classical intersecting surfaces of the ND
 sector.
 We demonstrated this for the simplest case of the Lorenz system with a linear
 dissipation.

\end{document}